\documentclass[conference]{IEEEtran}
\usepackage{multirow}
\usepackage{amsmath,amsfonts,amssymb}
\usepackage{mathptmx}
\usepackage{booktabs}
\usepackage{color}
\usepackage{cite}
\usepackage{balance}
\ifCLASSINFOpdf
\usepackage[pdftex]{graphicx}
\graphicspath{{figure/}}
\DeclareGraphicsExtensions{.pdf,.jpeg,.png}
\else
\fi
\ifCLASSOPTIONcompsoc
 \usepackage[caption=false,font=normalsize,labelfont=sf,textfont=sf]{subfig}
\else
  \usepackage[caption=false,font=footnotesize]{subfig}
\fi

\makeatletter
\def\blfootnote{\xdef\@thefnmark{}\@footnotetext}
\makeatother

\begin{document}

\title{High-quality Low-dose CT Reconstruction \\Using Convolutional Neural Networks with \\Spatial and Channel Squeeze and Excitation}

\author{\IEEEauthorblockN{Jingfeng Lu\IEEEauthorrefmark{2}\IEEEauthorrefmark{1}, Shuo Wang\IEEEauthorrefmark{3}\IEEEauthorrefmark{1}, Ping Li\IEEEauthorrefmark{4}, Dong Ye\IEEEauthorrefmark{3}}

\IEEEauthorblockA{\IEEEauthorrefmark{2}Metislab, School of Instrumentation Science and Engineering, Harbin Institute of Technology, Harbin, China\\Email: jingfeng.lu@hit.edu.cn
}
\IEEEauthorblockA{\IEEEauthorrefmark{3}School of Instrumentation Science and Engineering, Harbin Institute of Technology, Harbin, China\\Email:15B901018@hit.edu.cn, dongye@hit.edu.cn
}
\IEEEauthorblockA{\IEEEauthorrefmark{4}The Second Affiliated Hospital of Harbin Medical University, Harbin, China\\523371675@qq.com
}
}
\maketitle
\blfootnote{* indicates equal contribution}

\begin{abstract}
Low-dose computed tomography (CT) allows the reduction of radiation risk in clinical applications at the expense of image quality, which deteriorates the diagnosis accuracy of radiologists. In this work, we present a High-Quality Imaging network (HQINet) for the CT image reconstruction from Low-dose computed tomography (CT) acquisitions. HQINet was a convolutional encoder-decoder architecture, where the encoder was used to extract spatial and temporal information from three contiguous slices while the decoder was used to recover the spacial information of the middle slice. We provide experimental results on the real projection data from low-dose CT Image and Projection Data (LDCT-and-Projection-data), demonstrating that the proposed approach yielded a notable improvement of the performance in terms of image quality, with a rise of 5.5dB in terms of peak signal-to-noise ratio (PSNR) and 0.29 in terms of mutual information (MI).
\end{abstract}


\IEEEpeerreviewmaketitle

\section{Introduction} 

X-ray computed tomography (CT) imaging is one of the most common image modalities in clinics. As X-ray ionizing radiation leads to increased risk of radiation-related diseases, low-dose CT has raised a growing research interest in CT imaging for the reduction of radiation dose and related hazards. Nevertheless, low-dose CT typically introduces more noise than conventional CT, which potentially deteriorates the accuracy of diagnosis and requires more advanced algorithms to improve image quality. A variety of reconstruction approaches have been introduced to the CT imaging community. The existing studies can be categorized into sinogram filtration \cite{manduca2009projection}, iterative reconstruction \cite{xu2012low}, and post-processing technique \cite{chen2017low}.

In recent years, deep learning has emerged as the \textit{de facto} standard in various image processing problems. Inspired by the success of deep learning, many researchers have investigated deep learning methods for medical image reconstruction and achieved significant performance \cite{schlemper2017deep, hammernik2018learning, lee2018deep, lu2018unsupervised, lu2019fast, lu2020reconstruction}. Lee et al. \cite{lee2018deep} proposed a convolution neural network (CNN) for the reconstruction of magnetic resonance (MR) images from accelerated MR acquisition. In \cite{lu2020reconstruction}, Lu et al. proposed a multi-scale CNN to improve diverging wave (DW) ultrasound (US) imaging, yielding high-quality US images while using a small number of DW transmissions. In CT literature,  the systematic study of CNNs was first proposed in \cite{kang2017deep}, where a CNN using directional wavelets was demonstrated to be more efficient in removing noises induced from low-dose CT. Jin et al. \cite{jin2017deep} proposed to incorporate residual learning in the U-Net architecture, where large receptive fields were shown to be crucial to reduce globally distributed artifacts.

In this work, we propose a novel CNN architecture, the High-Quality Imaging network (HQINet), to reconstruct high-quality CT images from low-dose CT images. HQINet was a convolutional encoder-decoder framework, where the encoder was used to extract spatial information from three contiguous slices while the decoder was used to recover the spacial information of the middle slice. In particular, spatial and channel squeeze and excitation (scSE) modules were used to adaptively re-calibrate the feature maps. We provide experimental results evaluated on the low-dose CT Image and Projection Data (LDCT-and-Projection-data) \cite{mcCollough2020data} and show the effectiveness of the proposed method.

The remainder of this paper is organized as follows. Section II presents the proposed method. Section III provides the experiments and performance of the proposed method, which demonstrates the effectiveness of the proposed method.  Finally, we conclude this work in Section IV.

\begin{figure*}[!t]
\centering
\includegraphics[width=6in]{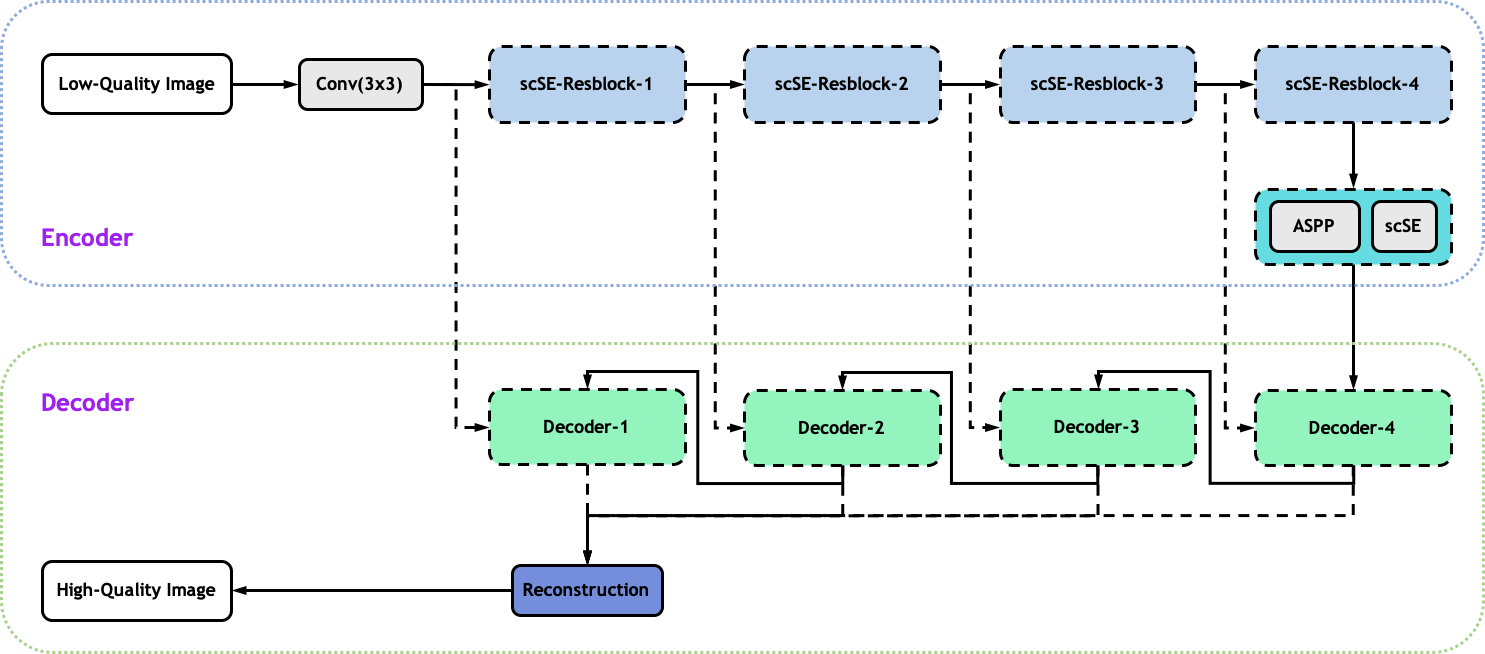}
\caption{Block diagram of the architecture of the proposed network.}
\label{f1}
\end{figure*}

\begin{figure*}[!t]
\centering
\subfloat[]{\includegraphics[width=0.8in]{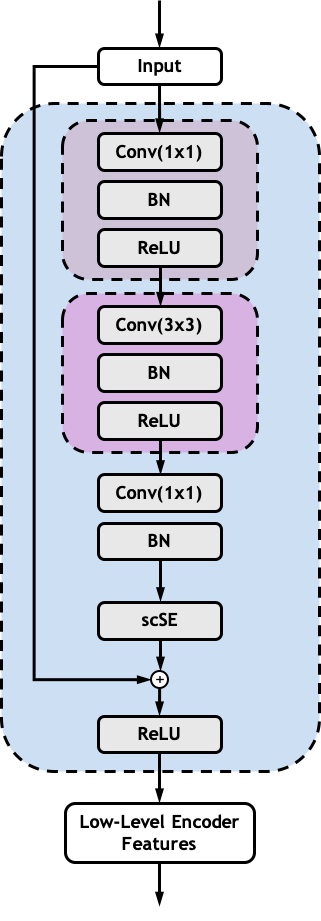}%
\label{a}}
\hfil
\subfloat[]{\includegraphics[width=3.6in]{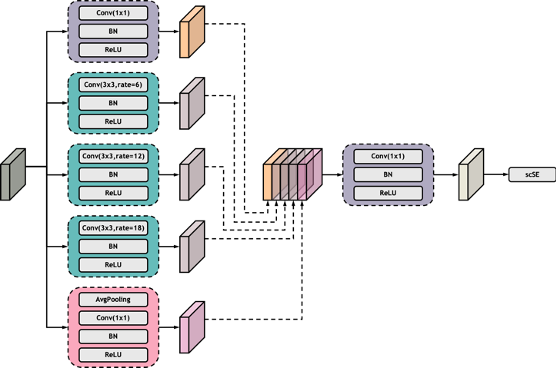}%
\label{b}}
\hfil
\subfloat[]{\includegraphics[width=1.4in]{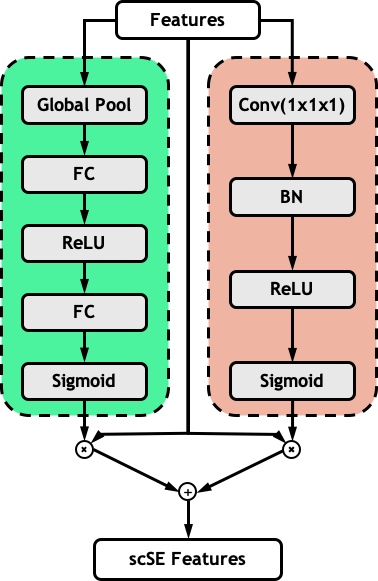}%
\label{c}}
\caption{Block diagram of the architecture of encoder blocks. (a) Spatial and Channel Squeeze and Channel Excitation (scSE) residual encoder  block; (b) Atrous Spatial Pyramid Pooling (ASPP) with depth separable convolution; (c) scSE block.}
\label{block}
\end{figure*}
	
\section{Methods}

Fig. \ref{f1} is a pictorial description of the proposed High-Quality Imaging network, which was an asymmetrical deep convolutional encoder-decoder architecture. The encoder network gradually reduced the resolution of feature maps and extracted high-level information, while the decoder network gradually recovered the spatial information. The inputs of HQINet were three successive low-dose CT slices, while the training reference was the full-dose CT image corresponding to the middle slice of inputs. Thus the HQINet was trained as a transformation model from low-dose images to full-dose images.

\subsection{Network Architecture}

The encoder network consisted of four residual blocks. We used a spatial and channel squeeze and excitation (scSE) module \cite{roy2018concurrent} at the end of each residual block \cite{He2016Deep}, as shown in Fig. \ref{a}. The scSE module (Fig. \ref{c}) was composed of spatial squeeze and channel excitation block (cSE) \cite{Hu2017Squeeze} and channel squeeze and spatial excitation block (sSE). The cSE module could adaptively tune the activations to emphasize the important channels and ignore less important ones, while the sSE module paid more attention to more relevant spatial locations and ignores irrelevant ones. The Atrous Spatial Pyramid Pooling (ASPP) module (Fig \ref{b}) \cite{Chen2017Rethinking} was used in the last residual block, which probed features at multiple scales by applying atrous convolution with different rates. To reduce the computation complexity while maintaining the similar performance of the ASPP module, the standard convolution  was replaced with atrous depthwise separable convolution \cite{Howard2017MobileNets,Zhang2017ShuffleNet,Chen2017Deeplab} in the ASPP module. The encoder network was constructed on the top of ResNet-50 \cite{He2016Deep}, while the first $7\times7$ convolution and $3\times3$ max pooling were replaced with $3\times3$ convolutions.

\begin{figure*}[!t]
\centering
\includegraphics[width=6.6in]{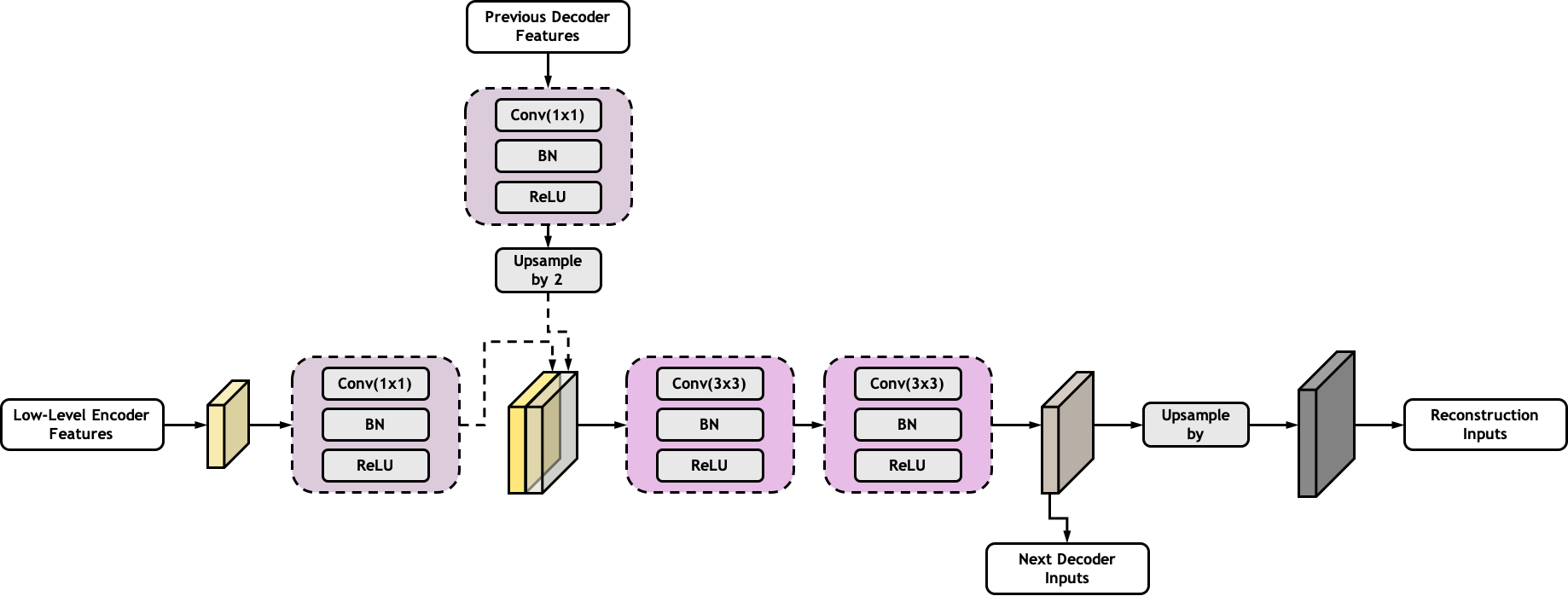}
\caption{Block diagram of the architecture of decoder block.}
\label{f3}
\end{figure*}

\begin{figure}[!t]
\centering
\includegraphics[width=3.4in]{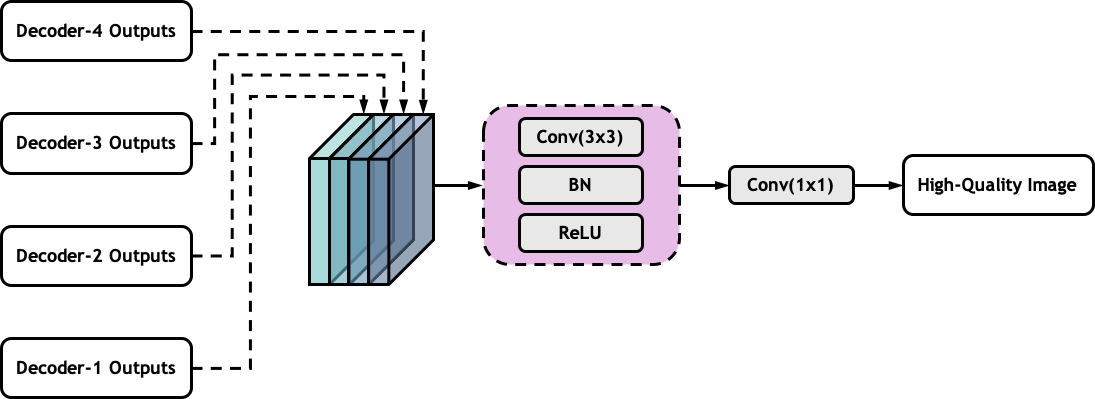}
\caption{Block diagram of the architecture of reconstruction block.}
\label{f4}
\end{figure}

The decoder network shared the same number of the decoder blocks as the encoder network. Each decoder block took the feature maps from the previous decoder block as well as the feature maps from the corresponding residual block. The architecture of the decoder block is illustrated in Fig. \ref{f3}. The decoder feature maps were first bilinearly $2\times$ upsampled and then concatenated with the corresponding encoder features maps. $1\times 1$ convolution was applied to the low-level encoder feature maps to lower the contribution by reducing the output channels. Two $3\times 3$ convolutions were then applied to concatenated maps to refine the features. The refined decoder features were used as the inputs of its next decoder block and bilinearly upsampled to the resolution of the full-dose CT images, which were used as the inputs of the reconstruction module.

The reconstruction module (Fig. \ref{f4}) took the feature maps from four decoder blocks to produce the final reconstruction. The feature maps of different scales were upsampled to the same resolution (the same as the references), and then concatenated as one group of feature maps. These features contained the texture information of different scales which were beneficial to recover high-quality images. $3\times3$ convolution was applied to fuse the features, while $1\times1$ convolution was used to reconstruct the final reconstruction.

\subsection{Loss Function}

The overall loss function $\mathcal{L}$ was
\begin{equation}
\label{loss}
\mathcal{L}=\alpha \mathcal{L}_{L1}+\beta \mathcal{L}_{SSIM},
\end{equation}
where $\mathcal{L}_{L1}$ and $\mathcal{L}_{SSIM}$ denoted L1 loss and structural similarity (SSIM) loss respectively, while $\alpha$ and $\beta$ was two constants for balancing the two losses which were empirically set to 0.85 and 0.15 respectively. 

First, the L1 loss$\mathcal{L}_{L1}$ was used. $\mathcal{L}_{L1}$ was the mean absolute error of two images.
\begin{equation}
\label{l1}
\mathcal{L}_{L1}=\parallel \hat I - I \parallel _1,
\end{equation}
where $\hat I$ and $I$ denoted reconstructed images and reference images respectively. 

Second, a perceptual-based SSIM loss was used. SSIM considered the interdependencies of object elements in the scene and modeled the distortion of the reconstruction into a combination of three different terms: luminance, contrast, and structure \cite{wang2004image}. Thus more relevant information of references was exploited with the SSIM loss. The SSIM loss was defined as: 
\begin{equation}
\label{ssim}
\mathcal{L}_{SSIM}=1- \frac{(2\mu _{\hat I} \mu _{I} + c_1)(2\sigma _{\hat{I}{I}} + c_2)}{(\mu _{\hat I}^2 + \mu _I^2 + c_1)(\sigma _{\hat I}^2 + \sigma _I^2 + c_2)},
\end{equation}
where $\mu _{\hat I}$  and $\mu _I$ were the average pixel intensities in $\hat I$ and $I$, $\sigma _{\hat I}^2$  and $\sigma _I^2$ were their variances, $\sigma _{\hat II}$ was the covariance between $\mu _{\hat I}$  and $\mu _I$, and $c_1$ and $c_2$ were two variables to stabilize the division.

\begin{figure*}[!t]
\centering
\subfloat[]{\includegraphics[width=1.8in]{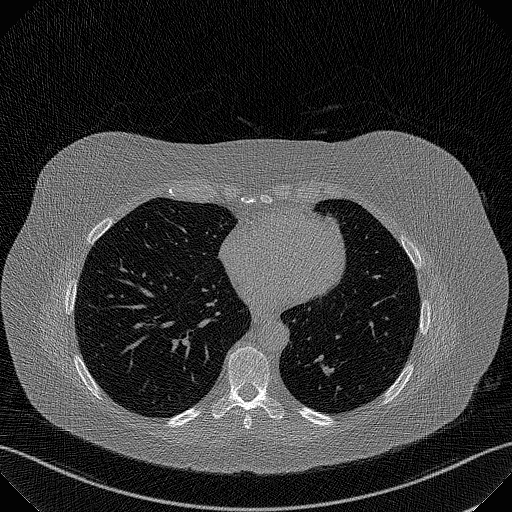}%
\label{low-1}}
\hfil
\subfloat[]{\includegraphics[width=1.8in]{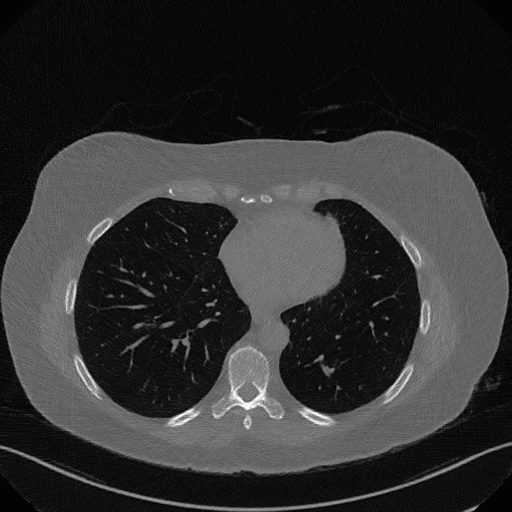}%
\label{prd-1}}
\hfil
\subfloat[]{\includegraphics[width=1.8in]{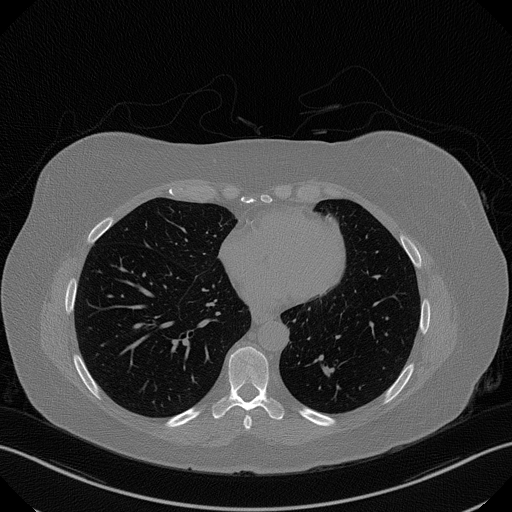}%
\label{ref-1}}
\hfil
\subfloat[]{\includegraphics[width=1.8in]{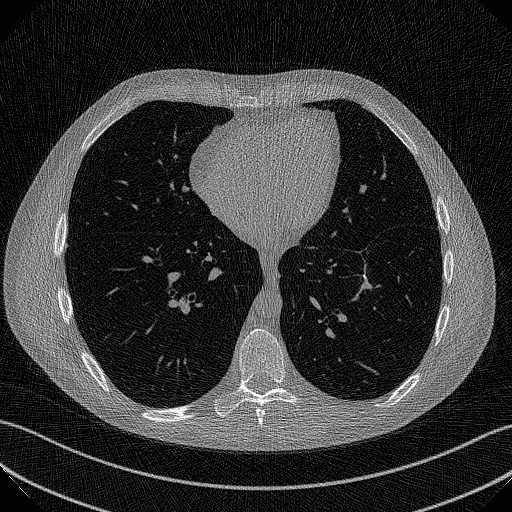}%
\label{low-2}}
\hfil
\subfloat[]{\includegraphics[width=1.8in]{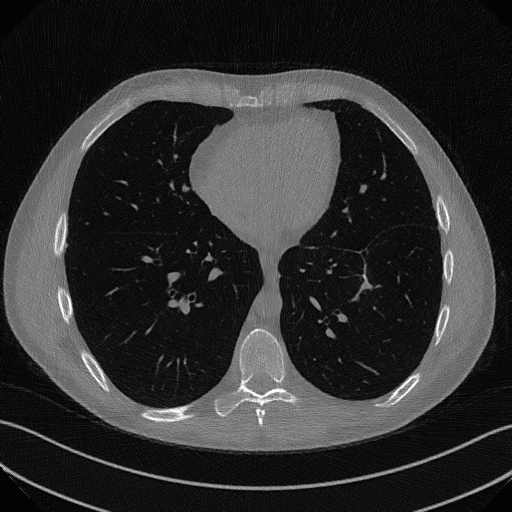}%
\label{prd-2}}
\hfil
\subfloat[]{\includegraphics[width=1.8in]{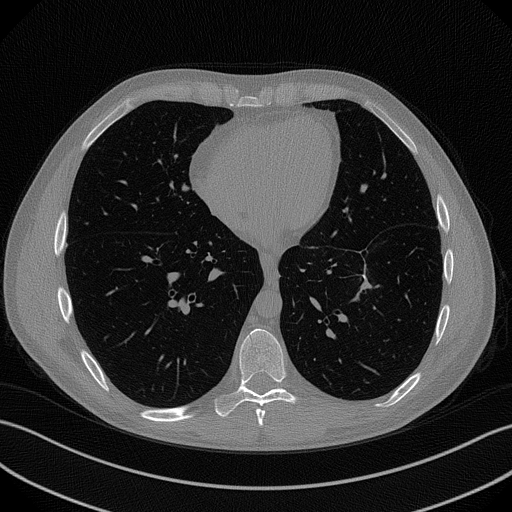}%
\label{ref-2}}
\caption{Examples images obtained from low-dose CT (network inputs), CNN reconstructions, and full-dose CT (references). Top to bottom: CT images from two patients samples, respectively. Left to right: low-dose CT images (a, d); CNN reconstructions (b, e) ; and full-dose CT images (c, f).}
\label{ct-image}
\end{figure*}


\subsection{Evaluation Metrics}

We used the four evaluation metrics: L1 error, normalized mean squared error (NMSE), peak signal-to-noise ratio (PSNR), and mutual Information (MI).

NMSE measured pixel-wise intensity differences between reconstructed and reference images.
\begin{equation}
\label{nmse}
{\rm NMSE}= \frac{\parallel \hat I - I \parallel _2^2}{\parallel I \parallel _2^2}.
\end{equation}

PSNR measured the degree to which image information rised above background noise.
\begin{equation}
\label{psnr}
{\rm PSNR} = 10 \log\frac{{\rm MAX}(I)^2}{\sqrt{{\rm MSE}(\hat I,I)}},
\end{equation}
where ${\rm MAX}(I)$ was the max pixel value in the reference image $I$, while ${\rm MSE} (\hat I,I)$ was the mean square error between $\hat I$ and $I$, defined as $\parallel \hat I - I \parallel _2^2$

MI measured mutual dependence between two images, which was calculated as
\begin{equation}
\label{mi}
{\rm MI} = \sum_{\hat{I},I}p_{\hat{I}I}(\hat{I},I)log\frac{p_{\hat{I}I}(\hat{I},I)}{p_{\hat I}(\hat{I})p_I(I)},
\end{equation}
where $p_{\hat{I}I}(\hat{I},I)$ was the joint distribution of $\hat I$ and $I$, and $p_{\hat I}(\hat I)$ and $p_I(I)$ were the marginal probability distribution of $\hat I$ and $I$.
\section{Experiments and Results}

A real clinical dataset: low-dose CT Image and Projection Data (LDCT-and-Projection-data) was used in the experiment. Specifically, samples acquired from 50 patients and 18 patients were randomly selected for training and evaluation respectively. In the training stage, the network weights were initialized using the truncated Gaussian distribution with zero mean and 0.01 standard deviation, while the biases were initialized to 0. The loss was optimized with gradient descent using the Adam optimizer \cite{kingma2014adam}, with a batch size of 88 and a fixed learning rate of 0.01. The training was implemented using Pytorch library on a Ubuntu server with 4 NVIDIA RTX TITAN GPUs, and converged in 20 epochs (about three hours).

In the testing phase, the images reconstructed by the optimal model in the training stage were compared to the low-dose and full-dose CT images. As displayed in Fig. \ref{ct-image}, the proposed model reconstructed high-quality images which were visually close to the reference (right column), and provided enhanced tissue structures compared to those from low-dose CT (left column). In particular, due to the decreased noises, the HQINet reconstruction showed an improved contrast between heterogeneous tissues, as compared to low-dose counterparts. To quantitatively assess the improvements, we report in Table \ref{table_test} the evaluation metrics described in Section III-C. The HQINet reconstruction showed a decrease of 0.02 and 0.05 in L1 error and NMSE respectively, and a gain of 5.5 dB in PSNR and 0.29 in MI respectively. 

\begin{table}[!t]
\renewcommand{\arraystretch}{1.25}
\setlength{\tabcolsep}{1.4mm}
\caption{Evaluation metrics of low-dose CT and CNN reconstruction.}
\label{table_test}
\centering
\begin{tabular}{c c c c c}
\toprule
model & L1 error & NMSE & PSNR [dB] & MI \\
\midrule
Low-dose & $0.043\pm0.020$ & $0.062\pm0.039$ & 25.81$\pm$5.86 & $1.01\pm$0.44 \\
HQINet & $0.021 \pm 0.006$ & $0.012\pm0.006$ & 31.30$\pm$2.69 & $1.30\pm$0.38 \\
\bottomrule
\end{tabular}
\end{table}

\section{Conclusion}

We present an approach based on a convolutional neural network (CNN) with spatial and channel squeeze and excitation (scSE) modules. The proposed approach aimed at learning a reconstruction operator to produce high-quality images using low-dose CT images. Evaluation on real clinical data demonstrated the effectiveness of the proposed method, with a notable improvement of the performance in terms of peak signal-to-noise ratio (PSNR) and mutual information (MI), and an image quality equivalent to that of high-quality full-dose CT images.


\ifCLASSOPTIONcaptionsoff
  \newpage
\fi
\balance
\bibliographystyle{IEEEtran}
\bibliography{Low_dose_CT.bib}

\end{document}